\documentclass{article}
\usepackage{graphicx}  % standard LaTeX graphics tool;
\usepackage{amsmath}   % For getting proper math eqns
\usepackage[compress]{cite}
\usepackage{amssymb}   % Used for getting various symbols eg. gtrsim, lesssim etc.
\usepackage{bm} % For bold math (esp of lower greek letters)
\usepackage{dcolumn}% Align table columns on decimal point
\usepackage{color}
\usepackage{mathrsfs}
\usepackage{amsfonts}
\usepackage{varioref}
\usepackage{textcomp}
\RequirePackage[colorlinks,citecolor=blue,urlcolor=magenta,linkcolor=blue]{hyperref}
\allowdisplaybreaks
\def\eq#1{{Eq.~(\ref{#1})}}

\def\cc{{cosmological\ constant}}

\def\dof{{degrees of freedom}}

\title{Microscopic origin of Einstein's field equations and the \textit{raison d'\^{e}tre} for a positive cosmological constant}

\author{T. Padmanabhan$^{1}$\footnote{Prof. T. Padmanabhan has passed away on 17th September, 2021 \cite{Bagla:2021tgl}, while this paper was under review in a journal.} and Sumanta Chakraborty$^{2}$\footnote{sumantac.physics@gmail.com}
\\
{$^{1}$\small{IUCAA, Post Bag 4, Ganeshkhind, Pune - 411 007, India}}\\
\small{$^{2}${School of Physical Sciences, Indian Association for the Cultivation of Science, Kolkata - 700032, India}}}

\date{\today}
\begin{document}

\maketitle
%%%%%%%%%%%%%%%%%%%%%%%%%%%%%%%%%%%%%%%%%%%%%%%%%%%%%%%%%%%%%%%%%%%%%%%%%%%%%%%%%%%%%%%%%%%%%%%%%%%
%%%%%%%%%%%%%%%%%%%%%%%%%%%%%%%%%%%%%%%%%%%%%%%%%%%%%%%%%%%%%%%%%%%%%%%%%%%%%%%%%%%%%%%%%%%%%%%%%%%
%%%%%%%%%%%%%%%%%%%%%%%%%%%%%%%%%%%%%%%%%%%%%%%%%%%%%%%%%%%%%%%%%%%%%%%%%%%%%%%%%%%%%%%%%%%%%%%%%%%
\begin{abstract}

In the paradigm of effective field theory, one  hierarchically obtains the effective action $\mathcal{A}_{\rm eff}[q, \cdots]$ for some low(er) energy degrees of freedom $q$, by integrating out the high(er) energy degrees of freedom $\xi$, in a path integral, based on an action $\mathcal{A}[q,\xi, \cdots]$. We show how one can integrate out a vector field $v^a$ in an action $\mathcal{A}[\Gamma,v,\cdots ]$ and obtain an effective action $\mathcal{A}_{\rm eff}[\Gamma, \cdots]$ which, on variation with respect to the connection $\Gamma$, leads to the Einstein's field equations and a metric compatible with the connection. The derivation \textit{predicts} a non-zero, positive, \cc, which arises as an integration constant. The Euclidean action $\mathcal{A}[\Gamma,v, \cdots]$, has an interpretation as the heat density of null surfaces, when translated into the Lorentzian spacetime. The  vector field $v^a$ can be interpreted as the Euclidean analogue of the microscopic degrees of freedom hosted by any null surface. Several implications of this approach are discussed.

\end{abstract}
%%%%%%%%%%%%%%%%%%%%%%%%%%%%%%%%%%%%%%%%%%%%%%%%%%%%%%%%%%%%%%%%%%%%%%%%%%%%%%%%%%%%%%%%%%%%%%%%%%%
%%%%%%%%%%%%%%%%%%%%%%%%%%%%%%%%%%%%%%%%%%%%%%%%%%%%%%%%%%%%%%%%%%%%%%%%%%%%%%%%%%%%%%%%%%%%%%%%%%%
%%%%%%%%%%%%%%%%%%%%%%%%%%%%%%%%%%%%%%%%%%%%%%%%%%%%%%%%%%%%%%%%%%%%%%%%%%%%%%%%%%%%%%%%%%%%%%%%%%%
\maketitle
%%%%%%%%%%%%%%%%%%%%%%%%%%%%%%%%%%%%%%%%%%%%%%%%%%%%%%%%%%%%%%%%%%%%%%%%%%%%%%%%%%%%%%%%%%%%%%%%%%%
%%%%%%%%%%%%%%%%%%%%%%%%%%%%%%%%%%%%%%%%%%%%%%%%%%%%%%%%%%%%%%%%%%%%%%%%%%%%%%%%%%%%%%%%%%%%%%%%%%%
%%%%%%%%%%%%%%%%%%%%%%%%%%%%%%%%%%%%%%%%%%%%%%%%%%%%%%%%%%%%%%%%%%%%%%%%%%%%%%%%%%%%%%%%%%%%%%%%%%%
\section{Introduction}

At length scales much larger than the mean-free-path, the dynamics of a fluid can be described in terms of, say, the Navier-Stokes equation. But at scales comparable to (or, smaller than) the mean-free-path of the fluid, one requires a more microscopic description, say, in terms of the many-body Schr\"{o}dinger equation governing  the dynamics of the atoms of the fluid. In this sense, the long wavelength behaviour of the fluid is described in a language which is quite different from that used in a more exact description. This process is hierarchical; as one probes the system at smaller and smaller scales --- proceeding from the atoms of the fluid to the nucleons and to the quarks which make up the nucleons --  the dynamical equations governing the relevant \dof\  undergo  transmutation. 

The hierarchical emergence  of such a sequence of effective descriptions of a system can be summarized, in the spirit of the Renormalization Group paradigm, along the following lines. We consider an ordered sequence of energy scales $E_1\ll E_2\ll E_3\cdots \ll E_n\ll E_{n+1}\cdots$   or, equivalently, length scales $\lambda_1 \gg \lambda_2 \gg \lambda_3 \cdots \gg \lambda_n \gg \lambda_{n+1} \cdots $ for the system. Coarse-graining the dynamics over the length scale $L$ with $\lambda_n > L  > \lambda_{n+1}$  will lead to an effective description at level-$n$ in terms of variables appropriate for that level. This idea can be stated in the  path integral language by the --- somewhat symbolic --- equation, connecting levels $n$ and $(n+1)$:
%%%%%%%%%%%%%%%%%%%%%%%%%%%%%%%%%%%%%%%%%%%%%%%%%%%%%%%%%%%
\begin{equation}
%% eqn 1
\int \mathcal{D} \xi\ \exp(-\mathcal{A}^{(n+1)}[q,\xi]) = \exp (-\mathcal{A}^n_{\rm eff} [q])~.
 \label{eqn1}
\end{equation} 
%%%%%%%%%%%%%%%%%%%%%%%%%%%%%%%%%%%%%%%%%%%%%%%%%%%%%%%%%%%
On the left hand side, the action $\mathcal{A}^{(n+1)}[q,\xi]$ depends on two \textit{sets} of variables $\{q\}$ and $\{\xi\}$ at level $(n+1)$; the path integral over the set $\{\xi\}$ leads to an effective\footnote{This is, similar to what we do in the standard Renormalization Group, with levels specified by range of momenta or by ``fast'' and ``slow'' variables.} action $\mathcal{A}^n_{\rm eff}[q]$ for the variables $\{q\}$ which survives. Of course, in this case the degrees of freedom $\{q\}$  existed even at level $(n+1)$ and did \textit{not} ``emerge'' at level $n$. So this is an \textit{effective} low(er) energy description for the degrees of freedom $\{q\}$ and not an \textit{emergent} description at level $n$. In this work, we will be concerned with such an effective (rather than emergent) description of gravity. (This fact is also emphasized in our comment (9) in the last section.)

There is significant evidence to suggest that the description of spacetime in terms of a metric (which obeys Einstein's field equations) is analogous to the description of a fluid which obeys the equations of fluid dynamics, with the Planck length $L_P \equiv (G\hbar/c^3)^{1/2}$ playing the role of the mean-free-path. In the usual approaches to quantum gravity, one \textit{tacitly assumes} that there are just two levels of description: one above Planck energy ($E>E_{\rm P}$ with $E_{\rm P}=1/L_{\rm P}$), described by some pre-geometric \dof\ and the other at $E<E_{\rm P}$,  described by standard geometric description of spacetime. \textit{But, it is  certainly possible that the trans-Planckian description of spacetime itself has a hierarchy of energy scales and dynamical emergence.} That is, there could exist a hierarchy of super-Planckian energy scales ($E_{\rm P}\ll E_1\ll E_2, \dots)$  with the effective \dof\ and physical description changing as we cross each scale. Given our ignorance of trans-Planckian physics, we should remain open to this possibility and explore it. 

%%%%%%%%%%%%%%%%%%%%%%%%%%%%%%%%%%%%%%%%%%%%%%%%%%%%%%%%%%%%%%%%%%%%%%%%%%%%%%%%%%%%%%%%%%%%%%%%%%%
%%%%%%%%%%%%%%%%%%%%%%%%%%%%%%%%%%%%%%%%%%%%%%%%%%%%%%%%%%%%%%%%%%%%%%%%%%%%%%%%%%%%%%%%%%%%%%%%%%%
%%%%%%%%%%%%%%%%%%%%%%%%%%%%%%%%%%%%%%%%%%%%%%%%%%%%%%%%%%%%%%%%%%%%%%%%%%%%%%%%%%%%%%%%%%%%%%%%%%%
\section{Effective action for gravity} 

In that context, our primary concern will be the final transition to the level ($E\lesssim E_P$) in which the description in terms of metric, connection etc arises from the immediately previous level ($E_{\rm P}<E<E_1$), with the two levels connected by \eq{eqn1}. The right hand side of \eq{eqn1} should contain an effective action in terms of suitable geometrical and matter variables such that its variation leads to Einstein's field equations. In the left hand side we have the unknown microscopic degrees of freedom, $\{\xi\}$, which need to be integrated out. Our aim is to construct a suitable left hand side such that the resulting $\mathcal{A}_{\rm eff}$ leads to Einstein's field equations. The advantage of this approach is that we are not attempting a full description of trans-Planckian physics at one go; instead we assume there is a description available at some energy range ($E_{\rm P}<E<E_1$) which, through \eq{eqn1}, leads to classical gravity.  

Incredibly enough, there exists a very simple model which achieves this goal. The ingredients of this model are  well motivated by the two clues \cite{tpreview} we have about quantum gravity; viz. the (i) nature of the \cc\ problem and (ii) the thermodynamics of null surfaces. (See \ref{appa} for this correspondence and further motivation, which we will not elaborate in the main text.) We will first highlight the technical derivation of the result  and then comment on several conceptual ingredients.

We choose the dynamical variables to be the connection $\Gamma$ and a vector field $v^{a}$, such that the action $\mathcal{A}[\Gamma,v^{a}]$ appearing in the left hand side of \eq{eqn1} is quadratic in the vector field, $v^a$, yielding:
%%%%%%%%%%%%%%%%%%%%%%%%%%%%%%%%%%%%%%%%%%%%%%%%%%%%%%%%%%%
\begin{equation}
%%% eqn 2
\mathcal{A}\left[\Gamma,v^a\right] = \int dV \,  \left\{ L^{4}\left[\kappa^{-1}R_{(ab)}(\Gamma)-  \bar T_{ab}(g)\right] v^a v^b\right\}~;\qquad \bar T_{ab}\equiv T_{ab}-\frac{1}{2}g_{ab}\left(g^{cd}T_{cd}\right)~.
\label{eqn2}
\end{equation}
%%%%%%%%%%%%%%%%%%%%%%%%%%%%%%%%%%%%%%%%%%%%%%%%%%%%%%%%%%%
Here $dV$ is the dimensionless proper volume element of the relevant region, which we will take   to be $dV = \sqrt{g}~d^4x/L^4$. The $L$ is an arbitrary, constant, length scale which is  introduced  (by writing $1=L^4/L^4$) just to make various expressions dimensionless; $R_{ab}(\Gamma)$ is the Ricci tensor dependent only on the connection $\Gamma^{a}_{bc}$. We will assume that the connection is symmetric ($\Gamma^{a}_{bc} = \Gamma^{a}_{cb}$) for simplicity. $T^{ab} $ is a second rank symmetric tensor which may depend on the metric but not on the connection and $\kappa$ is a coupling constant with the dimensions of $\textrm{(length)}^{2}$ in natural units. (Eventually, comparison with low energy physics will allow us to identify $T_{ab} $ with the stress tensor of matter and $\kappa$ with $8\pi L_{\rm P}^{2}\equiv 8\pi  G_{\rm N}$, where $G_{\rm N}$ is the Newton's gravitational constant.) We will treat the connection $\Gamma^i_{jk}$ and the metric $g_{ab}$ to be independent variables in the spirit of the Palatini approach.\footnote{Note: (a) The measure $dV = \sqrt{g}~d^4x/L^4$ is the natural choice for integration domain in Euclidean 4-space, in which we will work. However, the analysis goes through unhindered even if we choose a sub-region like, for e.g., a lower dimensional  hypersurface embedded in the $d=4$ space, with appropriate proper volume. (b) It is possible to relax the  assumption of $\Gamma^{a}_{bc} = \Gamma^{a}_{cb}$ and construct a theory with torsion, which we will not attempt here. (c) Since we are not assuming at this stage that $\Gamma$'s are derived from a metric, the Ricci tensor need not be symmetric. However, the contraction with $v^a v^b$ selects out the symmetric part of $R_{ab}$. (d) The $T_{ab}$ for most sources of physical relevance is independent of the connection; this is also an assumption which is frequently made in the Palatini approach.} The space(time) is treated as a $C^{\infty}$, 4-dimensional Euclidean manifold with the metric, connection and the vector field defined as at least $C^{2}$, $C^{1}$ and $C^{2}$ functions respectively. (The connection and the metric are treated as independent a priori.) This ensures that the Riemann tensor is continuous over the manifold. Additionally, any maps between the manifold, its tangent space and the real vector space is assumed to be continuous and differentiable to arbitrary orders.

In this approach, the connection $\Gamma^{a}_{bc}$ and the vector field $v^a$ are dynamical and the metric is treated as an independent auxiliary background field which goes for a ride. The gravitational sector depends on the metric only through the  $\sqrt{g}$ factor in $dV$ and if we choose a gauge with $\sqrt{g}=1$ (which can always be done) this dependence will also go away. From the standard identity,
%%%%%%%%%%%%%%%%%%%%%%%%%%%%%%%%%%%%%%%%%%%%%%%%%%%%%%%%%%%
\begin{equation}
%%%eqn 3
R_{(ab)}(\Gamma) v^a v^b = P^{ab}_{cd}\, \nabla_a v^c \, \nabla_b v^d -  \nabla_{a}\left(P^{ab}_{cd}v^{c}\nabla_{b}v^{d} \right)~.
\label{eqn3}
\end{equation}
%%%%%%%%%%%%%%%%%%%%%%%%%%%%%%%%%%%%%%%%%%%%%%%%%%%%%%%%%%%
where $P^{ab}_{cd} \equiv(\delta^{a}_{c}\delta^{b}_{d}-\delta^{a}_{d}\delta^{b}_{c})$ is entirely built from the Kronecker deltas, we see that the term $R_{ab}v^av^b$ differs from a ``kinetic energy'' term for $v^a$, viz. $P^{ab}_{cd}\, \nabla_a v^c \, \nabla_b v^d$, only by a total divergence. Even then, we will use the form of the action in \eq{eqn2}, since the path integral is then easy to handle.

So the vector field $v^a$, which is a degree of freedom that appears at the first trans-Planckian level ($E_P<E<E_1$), couples to both matter and gravity thereby inducing an indirect coupling between them; when we integrate out this degree of freedom, we will obtain an effective coupling between geometry and matter. Writing the Lagrangian in \eq{eqn2} as $M_{ab} v^a v^b$, with $M_{ab}\equiv L^{4}(\kappa^{-1}R_{(ab)}-\bar T_{ab})$, the path integral can  be evaluated by using the standard result (see the \ref{appb} for further details):
%%%%%%%%%%%%%%%%%%%%%%%%%%%%%%%%%%%%%%%%%%%%%%%%%%%%%%%%%%%
\begin{align}\label{tppi}
\int \mathcal{D}v^{a}&~\exp\left[-\int dV\,  v^a (M_{ab}) v^b \right]
\nonumber
\\
&\propto \exp\left[-\frac{1}{2}\int dV~\ln \left(\textrm{det}(M_{ab})\right)\right]
\nonumber
\\
&\propto \exp\left(-\mathcal{A}_{\rm eff}[M_{ab}]\right)~.
\end{align} 
%%%%%%%%%%%%%%%%%%%%%%%%%%%%%%%%%%%%%%%%%%%%%%%%%%%%%%%%%%%
The convergence of the path integral requires the tensor $M_{ab}$, treated as a matrix, to have positive definite eigenvalues.\footnote {In hierarchical emergence, the physics at a higher energy scale ($E>E_1\gg E_{\rm P}$) should have ensured this; we will assume this is the case. See \ref{appb} for further details.} Therefore we obtain the effective action for gravitational degrees of freedom to be:
%%%%%%%%%%%%%%%%%%%%%%%%%%%%%%%%%%%%%%%%%%%%%%%%%%%%%%%%%%%
\begin{equation}
 %%%eqn 15 of ppaer A
 \mathcal{A}_{\rm eff}=\frac{1}{2}\int dV~\ln \left\{\textrm{det}\left[L^4 (\kappa^{-1}R_{(ab)}(\Gamma)- \bar T_{ab})\right]\right\}~.
 \label{logdetact}
\end{equation}
%%%%%%%%%%%%%%%%%%%%%%%%%%%%%%%%%%%%%%%%%%%%%%%%%%%%%%%%%%%
When $T_{ab} =0$, this action is very similar to the one in Eddington gravity \cite{eddington}, except for the $\sqrt{g}$ factor in $dV$ and the logarithmic dependence. In spite of these differences, the variation of $\mathcal{A}_{\rm eff}$ does lead to Einstein's field equations, see ref. \cite{edgrav}. Here we will just highlight the important aspects of the derivation.

The variation of $A_{\rm eff}$ with respect to the connection (with the conventional boundary condition $\delta \Gamma^{a}_{bc} =0$ at the boundary) will lead to the equation
 %%%%%%%%%%%%%%%%%%%%%%%%%%%%%%%%%%%%%%%%%%%%%%%%%%%%%%%%%%%
 \begin{equation}
%%% eqn 49 of paper
\nabla_{a}\left(\sqrt{g}\, N^{bc}\right)=0~,
\label{eqn49}
\end{equation}
%%%%%%%%%%%%%%%%%%%%%%%%%%%%%%%%%%%%%%%%%%%%%%%%%%%%%%%%%%%
where the tensor $N^{bc}$ is defined through the relation $N^{bc}M_{cd} = \delta^{b}_{d}$. (In matrix language, $N$ is the inverse of $M$.) Eq. (\ref{eqn49}) tells us that  the expression within the bracket is a symmetric second rank tensor density with zero covariant derivative. This allows us to identify $N^{ab}\equiv (1/\lambda)g^{ab}$ and hence $M_{ab} = \lambda g_{ab}$, where $\lambda$ is a \textit{positive, non-zero} constant because $M_{ab}$ is a matrix with positive eigenvalues. Using the definition of $M_{ab}$, we immediately find that, $R_{ab} - \kappa \bar T_{ab} = \Lambda g_{ab}$ with $\Lambda\equiv\kappa\lambda$, which is equivalent to the Einstein's equations with a \textit{positive} cosmological constant:
%%%%%%%%%%%%%%%%%%%%%%%%%%%%%%%%%%%%%%%%%%%%%%%%%%%%%%%%%%%
 \begin{equation}
 %%eqn 12 of ppaer A 
G_{ab}+\Lambda g_{ab}
=\kappa T_{ab}~. 
\label{Einstein_eq}
\end{equation}
%%%%%%%%%%%%%%%%%%%%%%%%%%%%%%%%%%%%%%%%%%%%%%%%%%%%%%%%%%%
The condition $N^{ab} \propto g^{ab}$ reduces \eq{eqn49} to the form $\nabla_{a}(\sqrt{g}g^{bc}) =0$ showing that the metric and the connection are compatible with each other. We thus recover the standard Einstein's theory. The Bianchi identity will now lead to $\nabla_{a}T^{a}_{b}=0$, which will produce the dynamics of the matter source in simple scenarios (``Spacetime tells matter how to move''). More detailed discussion on the matter field equations has been presented in the subsequent section.

The conclusion that $N^{ab} \propto g^{ab}$ --- where $g^{ab}$ is the metric \textit{already present} in the matter sector of the action --- relies in a subtle manner upon (a version of) the Principle of Equivalence. It can be shown that the equation $\nabla_a (\sqrt{g} N^{bc}) =0$ always leads to the existence of a metric $q_{bc}$ compatible with the connection in $\nabla$. This introduces two \textit{different} metrics ($q_{ab}, g_{ab}$) on the manifold, which is not compatible with the Principle of Equivalence, unless they are identified with each other upto a constant multiplicative factor. 

%%%%%%%%%%%%%%%%%%%%%%%%%%%%%%%%%%%%%%%%%%%%%%%%%%%%%%%%%%%%%%%%%%%%%%%%%%%%%%%%%%%%%%%%%%%%%%%%%%%
%%%%%%%%%%%%%%%%%%%%%%%%%%%%%%%%%%%%%%%%%%%%%%%%%%%%%%%%%%%%%%%%%%%%%%%%%%%%%%%%%%%%%%%%%%%%%%%%%%%
%%%%%%%%%%%%%%%%%%%%%%%%%%%%%%%%%%%%%%%%%%%%%%%%%%%%%%%%%%%%%%%%%%%%%%%%%%%%%%%%%%%%%%%%%%%%%%%%%%%
\section{Discussion and Implications} 

We shall now make several technical and conceptual comments on the result. (Some of the technical points mentioned here also arises in the context of inclusion of matter in Eddington gravity and are discussed in greater detail in ref. \cite{edgrav}.)
\vskip 2mm

(1) \textit{Our approach leading to the Einstein's equations predicts the existence of a non-zero, positive $\Lambda$}! That is, the variational principle based on $A_{\rm eff}$  \textit{demands} a positive $\Lambda$, which  is consistent with the current cosmological observations.   In contrast, the standard Einstein-Hilbert action imposes no constraints on the cosmological constant. It can be zero or non-zero with either sign. This suggests that, even in the standard, classical, gravity \textit{it may be nicer to derive the Einstein's equations from our $A_{\rm eff}$ in Eq.(5) than from the usual Hilbert action}. Another reason to prefer $A_{\rm eff}$ over the Einstein-Hilbert action is that the variational problem based on $A_{\rm eff}$ is well-posed as it stands. The one based on the Einstein-Hilbert action is not well-posed, because one has to fix not only the metric but also normal derivatives of the metric on the boundary surface, which, in general, will not be consistent with the Einstein's equations. This requires one to add appropriate boundary terms to the Einstein-Hilbert action, in order to ensure that it has a well-defined functional derivative. The structure of these boundary terms are non-trivial and depends on the nature of the boundary surface \cite{Parattu:2015gga,Parattu:2016trq,Chakraborty:2016yna}. For the action $A_{\rm eff}$, presented in Eq.(5), such problems do not arise. The variational problem  is well-posed and does \emph{not} require additional  boundary terms. 
\vskip 2mm

(2) The field equation, viz., \eq{eqn49}, which emerges directly from the variation  involves second derivatives of the connection. Its first integral is given by Einstein's equation which involves only the first derivatives of the connection. Hence, structurally, $\Lambda$ has the status of an integration constant, the value of which has to be determined by some other physical principle (like, for e.g., in \cite{tphp}). It has been stressed in previous literature \cite{tpcc}, that such an approach holds the best prospect for solving the \cc\ problem. It is gratifying that the our approach implements this requirement in a very natural fashion.
\vskip 2mm

(3) The effective action $A_{\rm eff}$ we have obtained is \textit{not} the Hilbert action\footnote{So if we start with the Hilbert action, to describe the quantization of gravity, we may not get anywhere, as is evident from decades of efforts!.}; all the same, the variation of $A_{\rm eff}$ with respect to the connection leads to the Einstein's equations. In this connection, the following aspect is worth highlighting. In the conventional approach to Einstein's theory, the total action is the sum of the Hilbert action $A_{\rm H}$ and the action for matter fields $A_{\rm m}$. The metric is present in both $A_{\rm H}$ and $A_{\rm m}$ and the variation of the total action with respect to the metric leads to the field equations. In the absence of matter, even in pure gravity, one can still vary $A_{\rm H}$ with respect to the metric and obtain the vacuum Einstein's equations. So the metric can be treated as a dynamical variable (which is to be varied in the action) even in the absence of matter. However, in the formalism developed here, the situation is subtly different. In the absence of matter, our action is \emph{independent} of the metric and hence is \emph{not} the dynamical degrees of freedom (which can be varied in the action) in the case of pure gravity. Thus the dynamical nature of the metric is somewhat diminished in our approach. The metric appears in the action \textit{only} in the presence of matter and, that too, as a background, non-dynamical field. This suggests that: (a) One should not possibly consider the metric as dynamical, since in the absence of matter the metric disappears altogether from the effective action. (This point of view --- viz., that the metric is not a dynamical variable and should not be varied in any action principle --- has been stressed in the previous literature on the emergent gravity paradigm \cite{tpcc,tpreview}.) (b) In the complete theory of quantum spacetime and matter, it is very likely that matter and metric will emerge together.  
\vskip 2mm

(4) We would like to emphasize that the field equation(s) for the matter degree(s) of freedom, in the present scenario, do \emph{not} follow from any variations of the effective action, presented in Eq. (\ref{logdetact}). This is because, the matter degree(s) of freedom are auxiliary and not dynamical in the effective action and hence they should \emph{not} be varied. Rather, after deriving the Einstein's equations from the variation of the effective action with respect to the connection, application of the Bianchi identity immediately yields, $\nabla_{a}T^{a}_{b}=0$. One can easily verify, along with scalar and electromagnetic fields, that the above condition indeed provides the desired matter field equations, with a few caveats. For example, the configuration $\phi=\textrm{constant}$ indeed satisfies $\nabla_{a}T^{a}_{b}=0$ for a massive scalar field, however this is not a solution of the corresponding field equation. Moreover, if more than one field is present, then one may need an extra assumption that $\nabla_{a}T^{a}_{b}$ vanishes for each one of these fields.

In the domain of semi-classical gravity, however, one must vary the matter degrees of freedom present in the action, since this will provide the saddle point configuration of the matter fields in the path integral approach. This is possible, even in the present context, following the technique of \cite{Padmanabhan:2007xy}. First of all, one needs to add the matter Lagrangian $L_{\rm matter}$ to Eq. (\ref{eqn2}), which will also translate in an additive manner to the effective action in Eq. (\ref{logdetact}). Subsequently, substitution of the classical gravitational solution in the effective action itself, will yield the $L_{\rm matter}$ as the only part of the effective Lagrangian dependent on the matter degrees of freedom. This is evident from Eq. (\ref{Einstein_eq}), as, $\textrm{det.}(R_{ab}-\kappa \bar{T}_{ab})=\Lambda^{4}\textrm{det.}(g_{ab})$, devoid of any matter degrees of freedom. Therefore, one obtains the matter field equations by variation of the on-shell effective action. From the above analysis, it is evident that the prescription presented in the previous section will provide the classical matter equations only in simple scenarios. For situations involving multiple classical fields, as well as quantum fields, extra prescriptions, as the ones discussed above, are necessary.

\vskip 2mm
(5) The approach presented here differs from the conventional modified gravity theories as well as from the phenomenological models of quantum gravity in two key aspects. First of all, the effective action derived here yields the Einstein's equations \emph{without} any additional terms or, additional degrees of freedom. While, all the modified gravity theories and phenomenological models provide some corrections over and above the Einstein's equations. Secondly, and possibly most importantly, the effective action presented here, arises from a more fundamental action, which has direct thermodynamic interpretation. This is crucial in order to understand the connection between the gravity and thermodynamics from a microscopic point of view, which other approaches, e.g., modified gravity theories and phenomenological models of quantum gravity do not exhibit. Therefore, the motivation and implications of the model presented here are different from other modified gravity models. 
\vskip 2mm

(6) A deeper insight into the nature of microscopic spacetime will require an understanding of the nature of the action for the vector field $v^a$, which was integrated over to obtain $A_{\rm eff}$. This, in turn, requires understanding the nature of the two terms --- the kinetic energy term and the total divergence term --- in the right hand side of  \eq{eqn3}. A strong clue about this is provided by  the study of the thermodynamics of null surfaces in the \textit{Lorentzian} spacetime, wherein one encounters exactly the same structure \cite{tpcc,nullsctp}. The  kinetic energy term acquires a direct physical meaning as a viscous dissipation rate of a Navier-Stokes fluid \cite{dns1,dns2,dns3} and the total divergence term leads to the difference in the heat content of the boundary surfaces. Similarly, the term $T_{ab} \ell^a\ell^b$  can be interpreted as the heating rate of the null surface due to matter crossing it, as seen by a local Rindler observer who perceives the null surface  as a local Rindler horizon. The combination $(T_{ab} - \kappa^{-1}R_{ab}) \ell^a\ell^b$ can then be interpreted as the total heating rate of the null surface and Einstein's equation arises as a heat balance equation. It is easy to see \cite{Padmanabhan:2007xy,tpcc} that extremising $(T_{ab}-\kappa^{-1} R_{ab} )\ell^a \ell^b$ with respect to  $\ell^a$ and demanding that the extremum holds for all null congruences, leads to Einstein's equations $G_{ab} + \Lambda g_{ab} = \kappa T_{ab}$, with the \cc\ arising as an integration constant. The \dof\ $v^a$ are the Euclidean analogue of the null vectors $\ell^a$ which occurs in the Lorentzian sector (see \ref{appa} for some more details). It is to be noted that the transformation of Lorentzian metrics to Euclidean ones is achieved by first going to the local inertial frame and then performing the Wick rotation, locally. As discussed in \ref{appa}, such a local Wick rotation is sufficient to establish the connection between points in the Euclidean spacetime with null surfaces in the Lorentzian spacetime and hence the thermodynamic interpretation naturally follows. This is the viewpoint we have adopted in this work as well.
\vskip 2mm

(7) The action in \eq{eqn2} is ultra-local in $v^a$ with no derivatives of $v^a$ appearing in the Lagrangian. At the next level in the hierarchy, as we go above the threshold energy $E_1\gg E_P$, this could change. For example, consider the Lagrangian $L\propto [\mu^2(v^a\square v_a) +M_{ab}v^av^b]$ where $\mu^2=\mathcal{O}(1/E_2^2)$. When $E_P<E< E_1\ll E_2$, the $\mu^2(v^a\square v_a)$ can be ignored and the Lagrangian reduces to the ultra-local, non-propagating, form $L\propto M_{ab}v^av^b$ we have used. But in the next level ($E_1<E\lesssim E_2$), this term $\mu^2(v^a\square v_a)$ will contribute and the $v^a$ will acquire the characteristics of a standard, propagating, vector field. Its field equation $\mu^2\square v^a=-M^{ab}v_b$ shows that it is sourced by $M_{ab}\propto (\kappa^{-1}R_{ab}- \bar T_{ab})$ which represent the departure from the heat balance equilibrium. (This is analogous to the thermal excitations in the null surface in the Lorentzian sector; see Eq.(\ref{thex}) of \ref{appa}.)
\vskip 2mm

(8) Another aspect of the vector field $v^a$ is worth mentioning: The path integral average $\langle v^a(x)v^b(x)\rangle$, using the action in \eq{eqn2} will be $\langle v^av^b\rangle \propto N^{ab}(\Gamma,g, T_{ab})$ in general, where $N$ is the inverse of $M$, when treated as matrices. But, \textit{on-shell}, this average is given by $\langle v^av^b\rangle\propto g^{ab}$, thereby providing a direct relation between the vector field and the metric. The interpretation of this result is worth pursuing.
\vskip 2mm

(9) In our calculation, we started with an action $\mathcal{A}[\Gamma, v, T]$, integrated out the \dof\ $v^a$, to obtain $\mathcal{A}_{\rm eff}[\Gamma, T]$. This requires the matter sector, represented by $T_{ab}$, to be already present at the previous level and go for ride when we integrate out $v^a$. A proper understanding of this requires knowledge about the description of matter sector at trans-Planckian energies, which we do not have at present. In our hierarchical approach, it is possible that the interpretation of $T_{ab}\ell^a\ell^b$, as the heat density of matter comes about from a previous level.  
\vskip 2mm

(10) We stress that we are \textit{not} working in the framework of emergent gravity paradigm or analog gravity paradigm (see e.g., \cite{analogue}) in this Letter. What we have obtained is an \textit{effective}  description of a theory at low energies, by integrating out some high energy degrees of freedom. This is similar to obtaining the four-Fermi interaction by integrating out gauge bosons in the electroweak theory. The low energy Fermionic modes do not ``emerge'' and were already present in the original theory as well; only the effective action describing their dynamics changes in the process. Similarly, we start with $(v^a,g_{ab},T_{ab})$, integrate out the vector field $v^a$ and obtain an effective coupling between gravity and the matter energy momentum tensor.

%%%%%%%%%%%%%%%%%%%%%%%%%%%%%%%%%%%%%%%%%%%%%%%%%%%%%%%%%%%%%%%%%%%%%%%%%%%%%%%%%%%%%%%%%%%%%%%%%%%
%%%%%%%%%%%%%%%%%%%%%%%%%%%%%%%%%%%%%%%%%%%%%%%%%%%%%%%%%%%%%%%%%%%%%%%%%%%%%%%%%%%%%%%%%%%%%%%%%%%
%%%%%%%%%%%%%%%%%%%%%%%%%%%%%%%%%%%%%%%%%%%%%%%%%%%%%%%%%%%%%%%%%%%%%%%%%%%%%%%%%%%%%%%%%%%%%%%%%%%
\section{Conclusion}

To summarise, the two most significant, new, features of this work are the following:

(a) Start with a system described by the Lagrangian $L(\Gamma,v)\propto [\kappa^{-1}R_{ab}(\Gamma)-T_{ab}]v^av^b$ containing two dynamical degrees of freedom $\Gamma^i_{kj}$ and $v^a$ and auxiliary (non-dynamical) fields $g_{ab},T_{ab}$. The vector field $v^a$ couples to $R_{ab}$ and to $T_{ab}$; integrating out $v^a$ leads to an effective action $\mathcal{A}_{\rm eff}(\Gamma)$ (see \eq{logdetact} above) coupling $R_{ab}$ to $T_{ab}$. The variation of this action with respect to $\Gamma$ leads to the Einstein's field equations with a positive definite \cc. This result, which arises from straightforward algebra, is of importance by itself. \textit{It provides a better alternative to the Hilbert action to derive Einstein's theory and explains why we see a positive \cc\ in Nature.}

(b) One key aspect of this work is a paradigm shift: We know that, at low energies ($E<E_P$) there are several energy/mass scales (for which we have no explanation) with new \dof\ coming into play as we go to higher and higher energies, in a hierarchical fashion.  \textit{There is absolutely no reason to believe that the situation at super-Planck scales $E>E_P$ should be any different.}  We recognize the fact that, there could be a hierarchy of super-Planckian energy scales $E_P\ll E_1\ll E_2\cdots$ with the effective description and the \dof\ changing as we cross each of these scales. The primary interest will then be  the transition to geometric description at  $E\lesssim E_P$ from the \textit{immediately preceding} scale $E_P<E<E_1$, with the dynamics related by \eq{eqn1}. We find that this transition can be adequately modeled by introducing a vector field $v^a$ in the regime $E_P<E<E_1$, which couples to both matter and gravity. On integrating out $v^a$ we obtain the effective action for gravity coupled to matter for the range $E<E_P$.

%%%%%%%%%%%%%%%%%%%%%%%%%%%%%%%%%%%%%%%%%%%%%%%%%%%%%%%%%%%%%%%%%%%%%%%%%%%%%%%%%%%%%%%%%%%%%%%%%%%
%%%%%%%%%%%%%%%%%%%%%%%%%%%%%%%%%%%%%%%%%%%%%%%%%%%%%%%%%%%%%%%%%%%%%%%%%%%%%%%%%%%%%%%%%%%%%%%%%%%
\section*{Acknowledgements}

Research of S.C. is funded by the INSPIRE Faculty fellowship from DST, Government of India (Reg. No. DST/INSPIRE/04/2018/000893) and by the Start-Up Research Grant from SERB, DST, Government of India (Reg. No. SRG/2020/000409). The research of T.P is partially supported by the J.C.Bose Fellowship of the Department of Science and Technology, Government of India. We thank Krishnamohan Parattu and Karthik Rajeev for comments on the draft.
%%%%%%%%%%%%%%%%%%%%%%%%%%%%%%%%%%%%%%%%%%%%%%%%%%%%%%%%%%%%%%%%%%%%%%%%%%%%%%%%%%%%%%%%%%%%%%%%%%%
%%%%%%%%%%%%%%%%%%%%%%%%%%%%%%%%%%%%%%%%%%%%%%%%%%%%%%%%%%%%%%%%%%%%%%%%%%%%%%%%%%%%%%%%%%%%%%%%%%%
\appendix
\labelformat{section}{Appendix #1} 
\labelformat{subsection}{Appendix #1}
%%%%%%%%%%%%%%%%%%%%%%%%%%%%%%%%%%%%%%%%%%%%%%%%%%%%%%%%%%%%%%%%%%%%%%%%%%%%%%%%%%%%%%%%%%%%%%%%%%%
%%%%%%%%%%%%%%%%%%%%%%%%%%%%%%%%%%%%%%%%%%%%%%%%%%%%%%%%%%%%%%%%%%%%%%%%%%%%%%%%%%%%%%%%%%%%%%%%%%%
%%%%%%%%%%%%%%%%%%%%%%%%%%%%%%%%%%%%%%%%%%%%%%%%%%%%%%%%%%%%%%%%%%%%%%%%%%%%%%%%%%%%%%%%%%%%%%%%%%%
\section{Emergent gravity paradigm and the  thermodynamics of null surfaces: succinct summary}\label{appa}

The emergent gravity paradigm attributes to the gravitational field equation the same conceptual status as the equations of, say, fluid dynamics or elasticity. Gravity is treated as emergent in this specific sense; we do not consider spacetime or other geometrical structures as emergent at this stage, though it may be necessary to do so at the next level of hierarchy. The mathematical structure underlying this approach is governed and guided by the following physical principles \cite{tpreview,tpcc}. 

\begin{enumerate}

 \item We know that the field equation for the matter sector remain invariant under the addition of a constant to the matter Lagrangian. In the conventional approach, gravity \textit{breaks} this symmetry. To solve the \cc\ problem, it is necessary \cite{tpcc} to restore this symmetry to the gravitational sector; that is, the gravitational field equations must be invariant under the transformation $T^a_b \to T^a_b + (\text{constant}) \, \delta^a_b$. This is not possible in any generally covariant variational principle based on a local Lagrangian if the metric is varied in an unrestricted form. 
 
 \item The thermodynamics of null surfaces suggests that gravitational field equations must have a thermodynamic interpretation \cite{tpcc,nullsctp,dns1,dns2,dns3}. We stress that \textit{it is not sufficient} to just derive an equation like $G_{ab} = \kappa T_{ab} $ by some kind of thermodynamic reasoning. As it stands, this equation equates apples (geometry) to oranges (matter stress tensor). It is necessary to reinterpret \textit{both} the sides of the above field equations for gravity in a thermodynamic language. 
 
\end{enumerate}

 These  principles can be implemented along the following lines. Consider any null surface in the spacetime generated by a null congruence $\ell^a(x)$. Then $T_{ab} \ell^a\ell^b$ can be interpreted as the rate of heat generation (viz., $dQ/\sqrt{\gamma}d^2xd\lambda$) due to matter crossing the null surface. Similarly, $ -\kappa^{-1} R_{ab}\ell^a\ell^b$ can be interpreted as the rate of heat generation  due to geometry. Therefore, the equation $ (T_{ab} - \kappa^{-1} R_{ab}) \ell^a\ell^b =0 $ represents the condition for zero net heat generation on a given null surface. This dynamical equation is clearly invariant under the transformation $T^a_b \to T^a_b + (\text{constant}) \, \delta^a_b$. Thereby implementing the principles in items (1)  above; the equation $\kappa^{-1}R_{ab}\ell^a\ell^b=T_{ab}\ell^a\ell^b$ also has clear thermodynamical interpretation for both sides, implementing item (2) above. 
 
 In fact, the field equation can be given a more explicit interpretation by using the identity in Eq.(3) in the main text, to rewrite $R_{ab}\ell^a\ell^b$. Consider a null surface generated by a conguence $\ell^a(x)$ parameterised to have constant surface gravity $\kappa\equiv 2\pi T$. Let the intersection $\mathcal{S}(\lambda)$ of the null surface with $t=$ constant hypersurfaces be 2-dimensional compact surfaces with metric $\gamma_{ab}$,  area element $\sqrt{\gamma}d^2x$ and area $A(\lambda)$ so that their entropy is $S(\lambda)\equiv A(\lambda)/4$. The proper volume measure on the null surface is then $\sqrt{\gamma}d^2xd\lambda$. We assume that the null surface evolves from and asymptotic past stationary state at $\lambda=\lambda_1$ (which could be at $\lambda_1=-\infty$) to another  asymptotic future stationary state at $\lambda=\lambda_2$ (which could be at $\lambda_1=+\infty$). Then  
 integrating $ (T_{ab} - \kappa^{-1} R_{ab}) \ell^a\ell^b =0 $ over a null surface with measure $\sqrt{\gamma}d^2xd\lambda$ will give 
 %%%%%%%%%%%%%%%%%%%%%%%%%%%%%%%%%%%%%%%%%%%%%%%%%%%%%%%%%%%
\begin{align}
\int_{\lambda_1}^{\lambda_2}&d\lambda\int_{\mathcal{S}(\lambda)}\sqrt{\gamma} d^2x\; T_{ab}\ell^a\ell^b\  
\nonumber
\\
&+\int_{\lambda_1}^{\lambda_2}d\lambda\int_{\mathcal{S}(\lambda)}\sqrt{\gamma} d^2x\; \mathcal{D} =T[S(\lambda_2)-S(\lambda_1)]~.
\label{heatbal}
\end{align} 
%%%%%%%%%%%%%%%%%%%%%%%%%%%%%%%%%%%%%%%%%%%%%%%%%%%%%%%%%%%
Here the two terms on the left hand side represent  the integrated heating rate of the null surface due to matter and gravity, with  
%%%%%%%%%%%%%%%%%%%%%%%%%%%%%%%%%%%%%%%%%%%%%%%%%%%%%%%%%%%
\begin{equation}
\mathcal{D}\equiv -\frac{1}{\kappa}P^{ab}_{cd}\left(\nabla_{a}\ell^{c}\right)\left(\nabla_{b}\ell^{d}\right)=2\eta \sigma_{ab}\sigma^{ab}+\zeta \Theta^{2}~,
\end{equation} 
%%%%%%%%%%%%%%%%%%%%%%%%%%%%%%%%%%%%%%%%%%%%%%%%%%%%%%%%%%%
being the viscous dissipation rate, where $\Theta\equiv d\ln\sqrt{\gamma}/d\lambda$ is the expansion and $\sigma_{ab}\equiv \gamma_a^c\gamma_b^d\nabla_c\ell_d-(1/2)\gamma_{ab}(\gamma^{cd}\nabla_{c}\ell_{d})$ is the shear; $\eta=(1/2\kappa),\zeta=-(1/2\kappa)$ are the shear and the bulk viscosity coefficients. This $\mathcal{D}$ is the same as the Navier-Stokes viscous dissipation term which occurs when Einstein's equations are recast as a fluid dynamical equation \cite{dns1,dns2,dns3}. (This term arises from the bulk term in the right hand side of the identity in Eq.(3) in the main text.) The right hand side of \eq{heatbal} is the difference in the heat contents of the two asymptotic states. (This arises from the total divergence term in the right hand side of the identity in Eq.(3) in the main text.) Demanding the validity of this heat balance equation on all null surfaces is equivalent to $G_{ab} + \Lambda g_{ab} = \kappa T_{ab} $ with the cosmological constant, $\Lambda$, arising as an integration constant.
 
The above situation corresponds to something akin to the thermodynamic equilibrium. As usual, equilibrium thermodynamics also contains information about the fluctuations around the equilibrium. A local Rindler observer coasting very close to a null surface will attribute to it a (limiting) temperature of the order of Planck temperature $T_P\equiv\beta_P^{-1}$. The Boltzmann factor, 
%%%%%%%%%%%%%%%%%%%%%%%%%%%%%%%%%%%%%%%%%%%%%%%%%%%%%%%%%%%
\begin{equation}
 \mathcal{P}[\ell^a] \propto \exp\left(-\beta_P |L_P^3(T_{ab}-\kappa^{-1} R_{ab}) \ell^a \ell^b|\right)~,
 \label{thex}
\end{equation} 
%%%%%%%%%%%%%%%%%%%%%%%%%%%%%%%%%%%%%%%%%%%%%%%%%%%%%%%%%%%
will then represent the probability that the degrees of freedom $\ell^a(x)$ are excited on the null surface.
 
As we have emphasized earlier, \eq{heatbal} can be obtained from a variational principle based on the Lagrangian $L_{\rm null} = (\kappa^{-1} R_{ab} - T_{ab}) \ell^a\ell^b$ by varying $\ell^a$ and demanding that the resulting equation holds for all null surfaces.
 In this Lagrangian, metric is just a background field and $\ell^a$ are the dynamical degrees of freedom residing on a null surface which are varied. These \dof\ couple both to matter and geometry thereby inducing an indirect coupling between matter and geometry. Matter curves geometry  because they both  couple to  the \dof\ on all null surfaces.

One can immediately see the structural similarity between this Lagrangian $L_{\rm null}$ and the one used in our Euclidean path integral (see Eq.(2) in the main text), obtained by replacing $\ell^{a}$ by $v^{a}$. In the Euclidean sector the vector field $v^a$ couples to both matter and gravity thereby inducing an indirect coupling between the two. When we integrate out this degree of freedom, we obtain an effective coupling between geometry and matter. 
 That is, the degrees of freedom in $v^a$ in the Euclidean space (which, of course, has no null surfaces) \textit{act as a proxy} for the null congruence $\ell^a(x)$ in the Lorentzian spacetime. 
 Since there are no  null surfaces or null vectors in the Euclidean sector, we cannot expect a strict algebraic equality between Euclidean vector $v^a$ and the null vector $\ell^a$ under analytic continuation. However, the correspondence between the two can be established in terms of a limiting process as follows:
 
 Consider an arbitrary, curved, Euclidean 4-space. Around a given point $\mathcal{P}$ we introduce a locally flat Riemann normal coordinates ($t_E, \bm{x}$) with $\mathcal{P}$ as the origin.   
 The equation $|\bm{x}|^2 + t_E^2 = \epsilon^2$ will then represent the surface of Euclidean sphere of radius $\epsilon$ and, in the limit of $\epsilon \to 0$, the equation 
 $|\bm{x}|^2 + t_E^2 = 0$ will represent a single point $\mathcal{P}$ (the origin) in the Euclidean space. On analytic continuation to Lorentzian sector, we get the corresponding equation $|\bm{x}|^2 - t^2 = \epsilon^2$ which will represent a hyperboloid in the Lorentzian space. The limit $\epsilon \to 0 $, leading to 
 $|\bm{x}|^2 - t^2 = 0$, will now represent a light cone emanating from the origin of the Lorentz space. \textit{In this sense, the collection of all points in the Euclidean space corresponds to the collection of all light cones in the Lorentzian sector.}
 
 A stretched horizon, say, at one Planck length away from the null surface  in the Lorentzian sector --- given by the hyperboloid $|\bm{x}|^2 - t^2 = L_{\rm P}^{2}$ ---  will have a normal vector $n_a \propto \nabla_a (|\bm{x}|^2 - t^2)$.  On analytic continuation, the hyperboloid will become a sphere in the Euclidean sector with radius equal to the Planck length: $|\bm{x}|^2 + t_E^2 = L_P^2$  with the normal $v_a \propto \nabla_a (|\bm{x}|^2 + t_E^2)$. In the limit of the stretched horizon coinciding with the null horizon, its normal will become a null vector $\ell^a$ in the Lorentzian sector, which of course has no (non-zero) counterpart in the Euclidean sector. However, the normals to the hyperboloids do have a one-to-one correspondence with the normals to the sphere in the Euclidean space. It is in this limiting sense that one can interpret the correspondence between the Lagrangian $L_{\rm null} = (\kappa^{-1} R_{(ab)} - T_{ab}) \ell^a\ell^b$ in the Lorentzian sector and $L_E = (\kappa^{-1} R_{(ab)} - T_{ab}) v^a  v^b$ in the Euclidean sector. 

%%%%%%%%%%%%%%%%%%%%%%%%%%%%%%%%%%%%%%%%%%%%%%%%%%%%%%%%%%%%%%%%%%%%%%%%%%%%%%%%%%%%%%%%%%%%%%%%%%%
%%%%%%%%%%%%%%%%%%%%%%%%%%%%%%%%%%%%%%%%%%%%%%%%%%%%%%%%%%%%%%%%%%%%%%%%%%%%%%%%%%%%%%%%%%%%%%%%%%%
\section{Evaluation of the path integral}\label{appb}

We give meaning to the path integral over the vector field $v^a$ by the usual procedure of discretising the space. We divide the relevant region of space into  cells of infinitesimal, dimensionless, proper volume $dV_n $ centered at the events $\mathcal{P}_1, \mathcal{P}_2,... \mathcal{P}_n, ... $, and replace the integrals of the form $f(x) dV$ by the sum over $n$ of $f(\mathcal{P}_n)$. Consider the integral, at the lattice center $\mathcal{P}_n$ labeled by $n$, of the form: 
%%%%%%%%%%%%%%%%%%%%%%%%%%%%%%%%%%%%%%%%%%%%%%%%%%%%%%%%%%%
\begin{align}
%% eqn 4
\int d^4 v_n &\, \exp\left[ - v^a (\mathcal{P}_n) M_{ab} (\mathcal{P}_n) v^b  (\mathcal{P}_n)\right]
\nonumber
\\
&\propto (\text{det}\, M(\mathcal{P}_n))^{-1/2}= \exp\left[- \frac{1}{2} \ln \, \text{det} \, M (\mathcal{P}_n)\right]~.
\label{eqn4}
\end{align} 
%%%%%%%%%%%%%%%%%%%%%%%%%%%%%%%%%%%%%%%%%%%%%%%%%%%%%%%%%%%
Multiplying the integrals over all lattice centers, we get, 
%%%%%%%%%%%%%%%%%%%%%%%%%%%%%%%%%%%%%%%%%%%%%%%%%%%%%%%%%%%
\begin{align}
%% eqn 5
\int \bigg( \prod_n d^4 v_n\bigg) &\exp \bigg[ - \sum_n v^a(\mathcal{P}_n) M_{ab} (\mathcal{P}_n) v^b (\mathcal{P}_n)\bigg] 
\nonumber
\\
&=\exp\bigg[- \frac{1}{2} \sum_n \ln  \text{det} \, M (\mathcal{P}_n)\bigg]~.
 \label{eqn5}
\end{align} 
%%%%%%%%%%%%%%%%%%%%%%%%%%%%%%%%%%%%%%%%%%%%%%%%%%%%%%%%%%%
In converting the proportionality in \eq{eqn4} to an equality, we have ignored a proportionality constant which is irrelevant to the variational principle, since it is independent of the dynamical variables. The product over all $n$ of the measure $d^4v_n$ goes over to the path integral measure $\mathcal{D} v^a$ in the continuum limit. The discrete sums over the events $\mathcal{P}_n$ becomes integral over the dimensionless proper volume $dV$, in the same limit, giving the result 
%%%%%%%%%%%%%%%%%%%%%%%%%%%%%%%%%%%%%%%%%%%%%%%%%%%%%%%%%%%
\begin{align}
%%% eqn 6
\int \mathcal{D} v^a &\exp\left[-\int dV\,  v^a (x) M_{ab}(x) v^b(x)\right]
\nonumber
\\
&=\exp\left[- \frac{1}{2}\int dV\, \ln  \text{det} \, M(x)\right]~.
\label{eqn6}
\end{align} 
%%%%%%%%%%%%%%%%%%%%%%%%%%%%%%%%%%%%%%%%%%%%%%%%%%%%%%%%%%%
This is, of course, a standard result in the quantum field theory where one usually writes ($\ln \text{det}\, M$) as ($\text{Tr}\, \ln M$) because $M$ is usually an operator. In our case, it is better to work with ($\ln  \text{det}\, M$) form of the path integral. 

For the sake of completeness, we will note the following technical points which are again handled in a conventional manner. 

(a) The normalization constant in \eq{eqn4} and \eq{eqn5} can be handled by taking the ratio of two path integrals, one with the matrix $M_{ab}(x)$ which we are interested in and the other with any matrix $C_{ab}$ with, say, constant entries. This will replace  $\ln \text{det}\, M$  in \eq{eqn5} and \eq{eqn6} by  $\ln ( \text{det}\, M / \text{det}\, C)$. Since $(\ln \text{det}\, C)$ is independent of the dynamical variables, e.g., in the present context it may be constructed out of the metric tensor, it is irrelevant for the variational principle. Taking such a ratio of two path integrals also ensures the scalar behaviour of the Lagrangian under coordinate transformations. 

(b) The result of the Gaussian integral in \eq{eqn4} assumes that the eigenvalues of the matrix $M_{ab}$ are positive definite for the convergence of the integral. This is indeed the most natural context and we have assumed that all the eigenvalues of $M_{ab}$ \emph{are} indeed positive definite. As we saw in the main text, this leads to the prediction that the cosmological constant is positive definite. It is, however, possible to give meaning to the path integral, even when the matrix has zero or negative eigenvalues, \emph{but only by introducing extra prescriptions}. (We mention this procedure purely for the sake of technical completeness.) The zero eigenvalues are simply excluded in computing the determinant. The integration over  $v_n$ corresponding to the factor with negative eigenvalue $\lambda_n = - |\lambda_n|$, giving rise to the integrand $\exp(-\lambda_n v_n^2) = \exp(|\lambda_n| v_n^2)$, is given meaning by rotating the contour of $v_n$ integration into complex line along the imaginary axis. (This is exactly how the integration over the conformal degree of freedom --- which has the ``wrong sign'' --- is  handled in Euclidean quantum gravity \cite{hawking}.) More simply, we can add to $M_{ab}$ another matrix $\epsilon Q_{ab}$ thereby making all the eigenvalues positive; after the computation we can take the limit of  $\epsilon \to 0$. This will lead to $\text{det}\, M$ possibly becoming negative thereby leading to an additive factor involving $\ln (-1)$ in the action; again, it does not affect the variational principle. Many of these prescriptions are frequently used while regularizing path integrals and we have spelt it out in detail only for the sake of completeness. Of course we prefer not to introduce additional prescriptions and simply demand  that $M_{ab}$ is positive definite, since it \textit{predicts} a \cc\ which is positive.

%%%%%%%%%%%%%%%%%%%%%%%%%%%%%%%%%%%%%%%%%%%%%%%%%%%%%%%%%%%%%%%%%%%%%%%%%%%%%%%%%%%%%%%%%%%%%%%%%%%
%%%%%%%%%%%%%%%%%%%%%%%%%%%%%%%%%%%%%%%%%%%%%%%%%%%%%%%%%%%%%%%%%%%%%%%%%%%%%%%%%%%%%%%%%%%%%%%%%%%
%%%%%%%%%%%%%%%%%%%%%%%%%%%%%%%%%%%%%%%%%%%%%%%%%%%%%%%%%%%%%%%%%%%%%%%%%%%%%%%%%%%%%%%%%%%%%%%%%%%
%Bibliography
\bibliography{References}

\bibliographystyle{./utphys1}
%%%%%%%%%%%%%%%%%%%%%%%%%%%%%%%%%%%%%%%%%%%%%%%%%%%%%%%%%%%%%%%%%%%%%%%%%%%%%%%%%%%%%%%%%%%%%%%%%%%
%%%%%%%%%%%%%%%%%%%%%%%%%%%%%%%%%%%%%%%%%%%%%%%%%%%%%%%%%%%%%%%%%%%%%%%%%%%%%%%%%%%%%%%%%%%%%%%%%%%
\end{document}